\documentclass{iau_FM}
\usepackage{graphicx}

\title[Analysis of astronomical data through sonification] 
{Analysis of astronomical data through sonification: reaching more inclusion for visual disable scientists}

\author[Johanna Casado, Wanda Diaz-Merced \& Beatriz Garc{\'i}a]   
{Johanna Casado$^1$$^2$,
Wanda Diaz-Merced$^3$$^4$
 \and Beatriz Garc{\'i}a$^1$$^5$}

\affiliation{
$^1$Instituto de Tecnolog\'{i}as en Detecci\'{o}n y Astropart\'{i}culas (CNEA, CONICET, UNSAM), Mendoza, Argentina. \\ email: {\tt johanna.casado@iteda.cnea.gov.ar} \\[\affilskip]
$^2$Instituto de Bioingenier\'{i}a, Facultad de Ingenier\'{i}a, Universidad de Mendoza, Argentina.\\
$^3$Office of Astronomy for Development (OAD - IAU),  Southafrica.\\
$^4$ Office of Astronomy for Outreach (OAO), Japan. \\
$^5$ Universidad Tecnol\'{o}gica Nacional, Argentina
}

\pubyear{2024}
\editors{Piero Benvenuti, ed.}
\begin{document}

\maketitle
\begin{abstract}
Most tools for astrophysical research was centered on visual display. Even after some studies shows that the use of sound could help the data analysis, and on the other hand
generate more accessibility. This fact motivates the creation of a tool centered on the researcher with and without visual impairments. To carry out this challenge, on this contribution, a theoretical framework based on visual impaired people was created and included on the sonoUno software. After that, the accessibility of the tool was analysed with the ISO standard 9241-171:2008.
\end{abstract}
\firstsection 
\section{Introduction}
Nowaday, most of the developments for data analysis are limited to visualization, although it has been shown that the use of sound as a complement to visualization improves the analysis of such data (\cite[Diaz-Merced, 2013]{Diaz-Merced_13}). In addition, it must be taken into account that the use of software limited to visual displays do not allow to visual impaired people perform their studies and jobs. This fact raises the question of why multimodal tools are not accepted.

At present, there are some projects that use sound as adjunct to visual displays, some of them are Cosmonic (http://rgb.iaa.es/cosmonic/), Sonification Sandbox (http://sonify.ps\\ych.gatech.edu/research/sonification\_sandbox/), MathTrak (https://prime.jsc.nasa.gov/\\mathtrax/) and xSonify (https://sourceforge.net/projects/xsonify/). The problem with the last three software is that they are not user centered. This feature is one of the basic guidelines for Human Computer Interaction (HCI) design \cite[(Stephanidis, 2001)]{Stephanidis_01}. According to \cite{Stephanidis_01} “The aim of HCI is to ensure the safety, utility, effectiveness, efficiency, accessibility and usability of interactive computer-based systems”. This is very important to ensure the use of an application. The ideal is to incorporate the user from the beginning of the development.

In this contribution we perform a theoretical framework based on literature review as a first approximation to a user centred graphical user interface (GUI) for a sonification software to analyse astronomical datasets.
\firstsection
\section{Overview}
With the crescent use of the informatic devices, the development of accessible and usable user interfaces is very important. According to \cite{Kavcic_05} some guidelines are a logical tab order, descriptive names for the interface elements, consistency across the interface and keyboard equivalence for all actions. A user case study \cite[(Mulliken, 2018)]{Muliken_18}, based on the web accessibility, highlight the need of consistency and a complete communication of the things happening on the screen.

Design a GUI for visual impaired users also implicate some usability requirements: task adequacy (must take into account that blind user communicate serially with the computer), dimensional trade-off (the 2D interface that not visually impaired perceive has to be similar to the 1D interface that visually impaired people perceive), behaviour equivalence (all the elements of the interface have to be accessible), semantic loss avoidance and device independency \cite[(Alonso et. al., 2008)]{Alonso_08}.

With the above recommendations in mind and a previous ISO standard analysis \cite[(Casado et.al., 2017)]{Casado_17}, the development of the sonoUno software (https://github.com/sono\\UnoTeam/sonoUno) was initiated. This tool is been programed on Python language, completely open source. After completing a functional application, the user interface needed to be more centred on the user, in this case a researcher with and without visual impairment.
\firstsection
\section{Implications}
\begin{figure}[b]
\begin{center}
\includegraphics[width=3.4in]{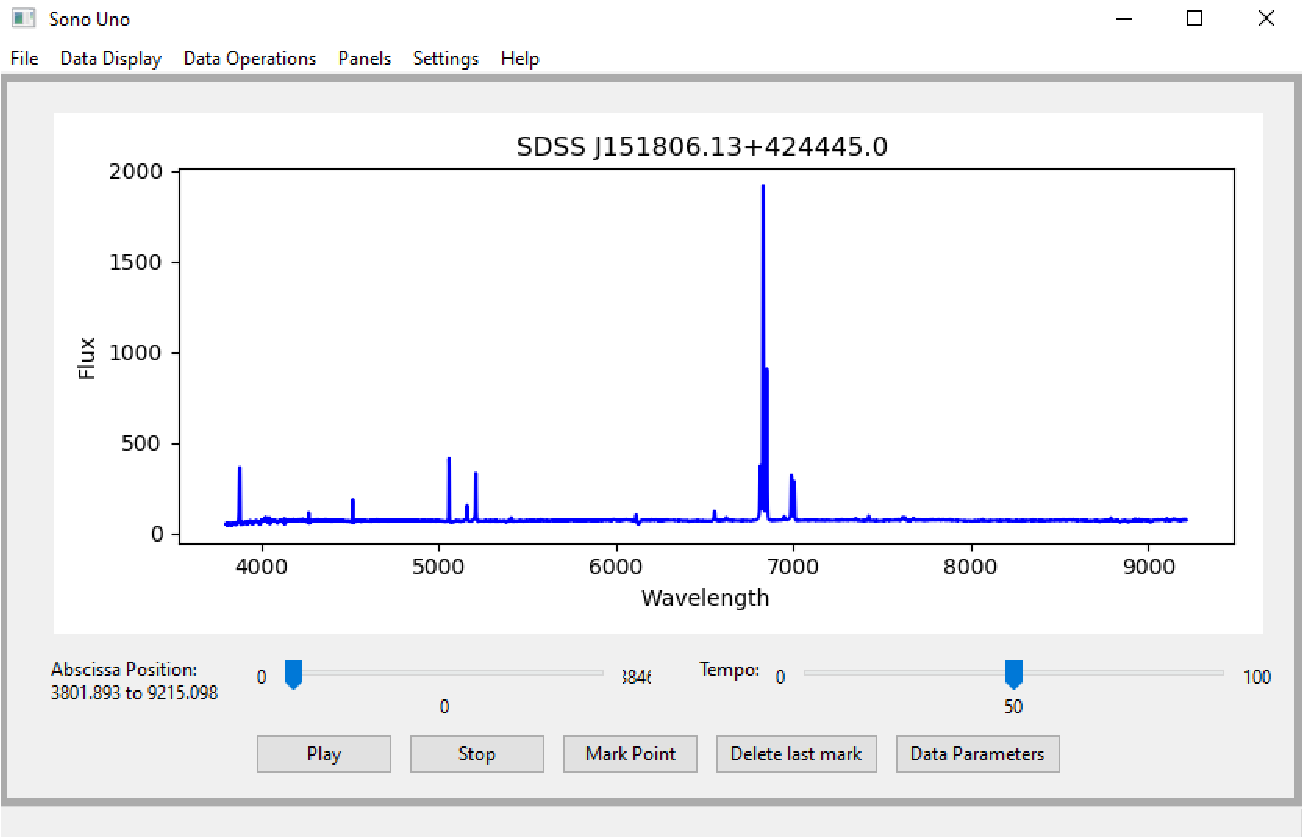} 
 \caption{First framework of the sonoUno software GUI. It contain the menu bar, the plot on the next row, abscissa position slider and tempo slider on the next row, and on the last row the play, stop, mark point, delete last mark and data parameters buttons. The data was extracted from https://www.sdss.org/}
   \label{fig1}
\end{center}
\end{figure}
\begin{figure}[b]
\begin{center}
\includegraphics[width=3.4in]{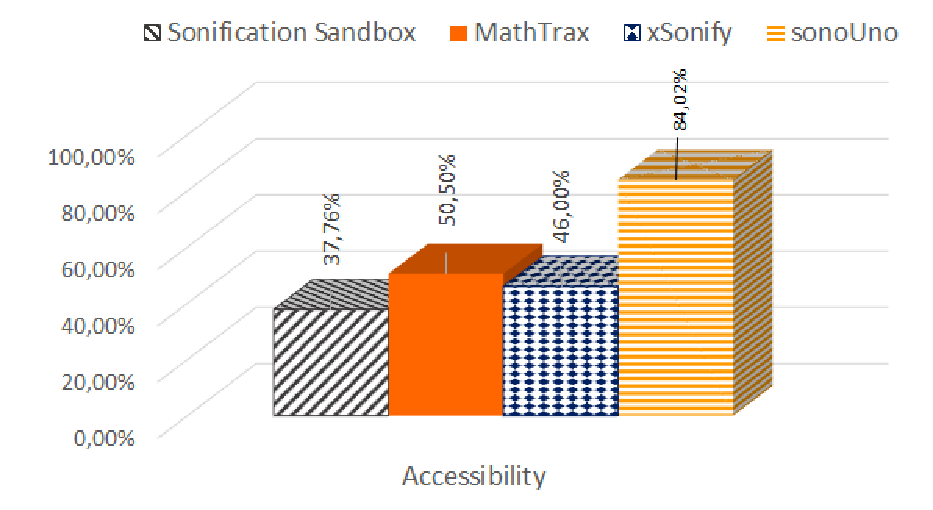} 
 \caption{Level of compliance of each software according to the ISO standard 9241-171:2008. The results are: Sonification Sandbox, 37.76\%; MathTrax, 50.50\%; xSonify, 46.00\%; and sonoUno, 84.02\%.}
   \label{fig2}
\end{center}
\end{figure}
Starting from an operative application with basic functionalities, a framework for the user interface was performed, based on attention mechanisms and coping strategies of people with visual impairment. With the information presented in the previous section and the authors expertice, some casual parameters was defined: there must be a linear relationship between the functionalities, the categorization of functionalities should not have overlays, hidden functionalities should not be presented and the interface should be simple.

With that in mind, an interface design improvement began and four categories were defined to placed all the functionalities based on the linear relationship between them (Table \ref{tab1}). This categories do not present overlapping, fulfilling one of the casual parameters.
\begin{table}
  \begin{center}
  \caption{Final approach of functionalities categorization of the user interface.}
  \label{tab1}
 {\scriptsize
  \begin{tabular}{|l|c|c|c|}\hline 
{\bf Data display} & {\bf Data operations} & {\bf Data configurations} & {\bf I/O options} \\ \hline
Abscissa position & Setting limits & Sound & Open \\ 
Tempo & Inverse & Plot & Save marks \\
Play & Square & Visual & Save sound \\
Pause & Logarithm & & Save plot \\
Mark point & Octave & & Quit \\
Stop & & & \\
\hline
  \end{tabular}
  }
 \end{center}
\vspace{-4mm}
\end{table}
The functionalities categorization match the first two casual parameters. To satisfy the next parameter all the functionalities are located in the menu bar, this characteristic sum to shortcut keys ensure that all the elements are accessible to the screen reader, the keyboard and the mouse. The tab navigation and the screen reader accessibility were tested on Mac (Voice Over on High Sierra distribution), Windows (NVDA on Windows 10) and Ubuntu (only the tab navigation on Ubuntu 16.04 and 18.04). The last item to comply is a simple interface, for that the design of the first framework of the GUI shows only the data display category (Figure \ref{fig1}). Then, the user can show or hide other panels with the other functionalities.

Once the improved version of GUI was implemented, the software accessibility was analysed according to the ISO standard 9241-171:2008 and the result was compared with the previous analysis of other software \cite[(Casado et.al., 2017)]{Casado_17} (Figure \ref{fig2}). The new analysis modify the previous result of the software Sonification Sandbox, MathTrax and xSonify, the items compliance is the same but the percentage change, during the new analysis the authors detected that seven items should not be taken into account because they refer to platform software. These items are 9.1.2-Allow parallel control of the functions of the pointing device using the keyboard, 9.3.12-Reserve accessibility keyboard shortcut assignments, 9.4.4-Allow changing the function assignment of the pointing device buttons, 9.4.13-Provide a means to find the pointer, 10.5.4-Allow to use "always ahead" windows (1st plane), 10.5.5-Allow the user to control multiple windows "always ahead" and 10.5.10-Allow windows to avoid focusing.

The new results did not consider this seven items from the previous analysis and amend the results of compliance. Even with these changes, the performance of sonoUno was around 30\% higher than the other softwares. The good compliance achieved answer to the user centered design executed from the beginning.
\firstsection
\section{Conclusions}
The sonoUno software present a high compliance to the ISO standard 9241-171:2008. This highlights the needs to think on the final user from the beginning of the development. The next step in this development, a focus group analysis, could improve the level of conformity according to this standard and deliver new user needs to meet by this tool.

This kind of exploration not only will bring people with disabilities to the field, but will also
increase the amount of scientific discoveries. One of our goals is to enhance the work of the scientists accepting different data exploration styles, more perspectives and experiences.
\firstsection

\end{document}